\documentclass[letterpaper]{article} 
\usepackage{aaai2027}  
\usepackage[hyphens]{url}  
\usepackage{graphicx} 
\usepackage{natbib}  
\usepackage{caption} 
\usepackage{algorithm}
\usepackage{algorithmic}

\usepackage{newfloat}
\usepackage{listings}
\DeclareCaptionStyle{ruled}{labelfont=normalfont,labelsep=colon,strut=off} 
\floatstyle{ruled}
\newfloat{listing}{tb}{lst}{}
\floatname{listing}{Listing}

\usepackage{booktabs}
\title{What Comes Next? Teens Learning About and Designing\\ Markov Chains Through Unplugged AI Activities}
\author {
    Daniel J. Noh\textsuperscript{\rm 1}\corresponding, 
    Luis Morales-Navarro\textsuperscript{\rm 2}, 
    Yasmin B. Kafai\textsuperscript{\rm 1}
}
\affiliations {
    \textsuperscript{\rm 1}University of Pennsylvania\\
    \textsuperscript{\rm 2}Aarhus University\\
    dnoh@upenn.edu, luismn@cc.au.dk, kafai@upenn.edu
}

\begin{document}

\maketitle

\begin{abstract}
A Markov chain is a process that describes the probability of a future event based on the most recent event or sequence of events. Despite the foundational role of Markov chains in probabilistic modeling, now particularly relevant given the heightened attention towards generative language models, few efforts have addressed how young people can learn about them. Building on traditions of developing unplugged activities for computing education, we present a collection of five unplugged activities that introduce teens to Markov chains as a simplified entry point for reasoning how generative language models produce text. These activities engage youth with foundational concepts, including state-based prediction, tokenization, n-grams, and state transition diagrams, and in building their own paper-based Markov chains using self-collected datasets using modified, familiar games like rock-paper-scissors. In this experience report, we present examples of student work and reflections from a pilot workshop with 19 consenting teenagers (ages 14-15) who completed these activities before building their own generative language models. We observed that the activities helped participants reason about text generation as a probabilistic process, use patterns in data to make predictions, and explain how training data shapes model output. We hope these activities can inspire future efforts to develop activities that connect foundational methods to the technologies that increasingly shape young people’s everyday lives.
\end{abstract}

\begin{links}
    \link{Activities}{https://daninoh.github.io/projects/markov-chains}
\end{links}

\section{Introduction}

Over the past years, as generative AI systems have become more accessible and pervasive in youths’ lives, there has been growing interest in supporting young people in understanding how these systems work. While generative AI tools have been rapidly introduced into K-12 learning contexts \citep{10.1145/3702652.3744217}, less research has explored how young learners understand the underlying mechanisms of these technologies. \citet{Smyslova02112025} note that, despite the need to help youth navigate the capabilities and limitations of algorithmic systems, there is a lack of empirical studies investigating how youth engage with and make sense of the functional inner workings of generative AI systems. Rather than positioning youth as naive users of blackboxed systems, learning opportunities should help demystify how models make predictions, encouraging learners to reflect on their design. 

Further, most learning activities assume access to computers, tablets, prior coding or data science knowledge, or familiarity with AI concepts \citep{10.1145/3641552, ali2024constructing, 10.1145/3773077.3806107, 10.1145/3773077.3812183} raising questions about who gets to engage in learning about the systems shaping youth’s lives and how these learning opportunities can be implemented. Unplugged approaches to AI/ML education, or activities that do not rely on computers, offer a flexible and accessible alternative for settings where devices and internet access may not be reliably available, such as museums, after-school programs, or under-resourced classrooms, and amid school district bans on devices and generative AI systems in the classroom across the world \citep{nyc2026mamdani, nytimes2026lausd, cnn2026denmark, reuters2026norway}. Additionally, these approaches also make it possible to teach core computing concepts without requiring prior coding or data science proficiency, lowering the barrier to entry for learners and facilitators alike. Thus, unplugged activities broaden access through their low cost, engage learners with little to no prior computing experience, and support easy collaboration among participants. 

Building on the traditions of developing unplugged activities for computing education \citep{bell1998computer} and on more recent efforts to develop unplugged activities for AI/ML learning \citep{10.1007/978-3-030-33759-9_10, 10.1145/3786761, 10.1145/3737609.3747112}, we developed a set of unplugged learning activities on predictive models using Markov chains. In this experience report, we present five sequential, paper-based activities that introduce youth to core AI/ML concepts, from probabilistic prediction (tokenization, n-grams, and state transition diagrams) to how the quantity and quality of training data shape a model's behavior, through games and materials that require no prior technical background other than an understanding of fractions. We present findings from a pilot workshop in which 14–15-year-old teens engaged with these activities before building their own generative language models (both Markov chain-based and GPT-based), and we describe illustrative observations of participants' reasoning. This paper contributes (1) a set of five unplugged activities in which learners calculate next-token probabilities themselves and construct transition tables and state transition diagrams from data they collect or design, and (2) implementation of lessons with illustrative observations suggesting that unplugged, paper-based engagement with Markov chains can help youth reason about the role of probability in generative language models.

\section{Background}
\subsection{Natural Language Processing for K–12}
With the advent of large-scale generative language models (LLMs), researchers have been investigating how to support learners in understanding these black-box AI/ML systems. These efforts have involved engaging teens in training small language models using the nanoGPT framework \citep{10.1145/3773077.3806107}, and building web-based tools that can support teens in training models on their own in the browser \citep{10.1145/3769994.3770061}. While such efforts are important in providing learners with opportunities to design systems, they still blackbox the transformer-based neural networks that power the models. Making such mechanisms more transparent requires advanced mathematical knowledge that is typically not taught in K-12 schools. As such, researchers have looked for ways to introduce young people to the idea of probabilistic models using simpler mechanisms \citep{10.1145/3737609.3747112, Smyslova02112025}.

One example of probabilistic models is Markov chains. Markov chains use state-based predictions to describe the probability of each possible event in a set of data. Notably, Markov chains are memoryless (Markov property), which means the probability of the next state depends only on the current state, not on the full sequence of past states \citep{dayIntroChains}. Learning about these models can serve as a simplified entry point for introducing learners to the probabilistic core of generative language models, which use transformer-based neural networks to generate more complex text \citep{Smyslova02112025}. As such, Markov chains offer one accessible way for students to begin reasoning about how generative AI produces text. In higher education, Markov chains have traditionally been taught in the context of linear algebra, requiring a working knowledge of matrix multiplication \citep{dayIntroChains}, but far fewer efforts have made these concepts accessible to K-12 learners. Yet, just with a basic understanding of fractions, young people can engage with Markov chains.

More recently, K-12 AI/ML education initiatives have begun using Markov chains to introduce the idea that popular large language models (LLMs) like ChatGPT are, at their core, statistically powered predictive language models \citep{codeorg2023chatbots}. As \citet{chen2024markov} note, ``Markov chains are much less powerful than GPT models, but they have the advantage of being easier to understand and simulate.'' Both Markov chains and LLMs generate text by predicting the next token from probabilities, but LLMs generate text based on far more context and are trained using neural networks \citep{Smyslova02112025}. Building on this idea, \citet{Smyslova02112025} used two different Markov chain-powered interactive text generation demos with high school students and found that, after learning about Markov chains in the context of a language arts class, they developed a stronger understanding of probabilistic thinking, greater awareness of the limitations of AI/ML systems, and an ability to engage in discussions of human versus AI authorship. The web-based interactive tools used in that study, however, were designed to be intentionally open-ended, which supports exploration but also requires extensive instructional scaffolding from a trained facilitator and access to a computer with an internet connection.

In the following section, we discuss prior work on, and the benefits of, unplugged activities for teaching AI/ML principles.

\subsection{Unplugged AI Activities}
Using unplugged activities to teach computing concepts is a long-standing tradition, with some of the earliest examples introducing children to fundamental CS ideas such as information theory, cryptography, and even the Turing test \citep{bell1998computer, bell2009computer}. \citet{10.1145/1539024.1508951} highlight that unplugged CS activities work well because they do not require computers and are generally based around a game or challenge, which can foster playfulness and motivation. They are also often student-directed, encouraging teamwork and communication, and are easy to prepare and implement across a range of contexts.

Within AI/ML specifically, this tradition dates back at least to the early 2000s, with van Halteren (\citeyear{10.3115/1118108.1118109}) describing teaching parsing algorithms for natural language processing through games and cards. As generative AI systems have emerged in children’s everyday experiences through popular LLMs such as ChatGPT, LLM-powered toys, voice assistants, and tutoring systems, there have been increasing efforts to develop unplugged activities for AI/ML learning contexts \citep{10.1007/978-3-030-33759-9_10, baldoni2026unplugged, 10.1145/3786761}. Unplugged AI activities, similar to unplugged CS activities, are particularly well-suited to this goal of designing accessible modes of learning about AI/ML, as these activities are generally low-cost and can be facilitated by caretakers and educators with little computing or AI expertise \citep{10.1145/3706599.3719876}. This makes them particularly valuable for time-limited experiences and for contexts where learners bring little or highly varied prior exposure to AI/ML, such as drop-in museum activities or informal after-school workshops.

Several prior efforts illustrate the range of AI/ML concepts that unplugged activities can address. Much of this work has targeted core technical ideas. \citet{10.1145/3545945.3569772} designed unplugged activities for teaching decision trees and k-means clustering. Likewise, \citet{pmlr-v207-michaeli23a} developed similar unplugged activities on decision trees, but included additional hands-on data mining in Orange (data mining/visualization software). Others have pushed into a broader set of concepts, with \citet{10.1145/3465074.3465078} addressing semantic networks and knowledge representation, and \citet{10.1007/978-3-030-33759-9_10} spanning classification, neural networks, reinforcement learning, and the Turing test. Other work has focused on the socio-ethical dimensions of AI. \citet{10.1145/3786761} developed a set of unplugged activities addressing data privacy, ethics, and the distinction between human and artificial intelligence, through which students demonstrated a growing awareness of AI’s capabilities and constraints, ethical implications of data collection, and understanding of data privacy and AI-driven personalization. \citet{10.1145/3392063.3394396} similarly created unplugged activities centered on AI ethics for middle school-aged children, engaging them in stakeholder analysis and a paper-based redesign of YouTube.

Regarding Markovian models, \citet{10.1609/aaai.v40i47.41515} developed a collection of unplugged activities built around probabilistic modeling. These activities center on Hidden Markov Models and Markov decision processes, which fall within the broader landscape of sequence modeling and reinforcement learning and require more complex mathematics, such as linear algebra. Indeed, the activities are designed for a ``university-level Introduction to AI course'' (p. 1) and are less connected to the generative mechanism of an LLM than the Markov chain.

Most relevant to the present work, a budding body of research has begun to unpack the ``black box'' of LLMs specifically, engaging children in the inner workings of text generation. \citet{10.1145/3773077.3806141}, for example, developed a collection of activities called The Engine Room that teach children about text generation through tangible bag-of-words models, probability-based text generation using wheel spinners, and tokenization. These activities were implemented in language arts classrooms, where teachers scaffolded discussions comparing human and AI-generated writing. \citet{10.1145/3786228.3786237} similarly has developed LLMs Unplugged, a collection of activities in which students build their own n-gram language models using cut-outs of existing children’s novels as training data. The activity prompts learners to determine the following token (or word) by ``pick[ing] from many candidates by eye.'' Of note, both The Engine Room and LLMs Unplugged do not engage with calculating probabilities or creating transition matrices.

These examples show that unplugged AI activities can be accessible in both content knowledge and required materials, while remaining playful and flexible enough to adapt across contexts. Our activities were designed with these same goals, extending prior unplugged AI/ML work specifically toward the mechanics of probabilistic modeling.

\section{Methods}
\subsection{Activity Design}
In this section, we provide activity descriptions for five scaffolded, unplugged activities that engage youth in concepts ranging from tokenization to dataset bias. These activities were designed to be sequential, with each activity building on concepts from the previous activity (see Table \ref{tab:activities}). They were also iteratively developed across three pilot workshops: first in a high school classroom, then with a group of teens at a science center, and finally with a second cohort of teens at the same science center. Each iteration surfaced lessons that shaped the current versions of the activities. For example, in the ``Rock, Paper, Scissors'' activity, several groups spontaneously engaged in multiple rounds of the game, with one group engaging in a ``best-of-three'' format; we incorporated this structure into the current version. Similarly, when completing the ``Guess the Emoji'' activity, which prompted learners to guess the single most probable final token given peer-designed datasets, participants completed the task in richer ways than expected. 

\begin{table}[t]
\centering
\begin{tabular}{p{3cm} p{4.5cm}}
\toprule
\textbf{Activities} & \textbf{Key Concepts} \\
\midrule
What Comes Next? & Prediction is based on prior data \\
\addlinespace
Pair of N-imals & More context can change prediction outcomes \\
\addlinespace
Sunny, Rainy, Cloudy & State transition diagrams, prediction models as a system \\
\addlinespace
Rock, Paper, Scissors & Data quantity and quality, bias, representativity \\
\addlinespace
Guess the Emoji & Dataset creation, testing \\
\bottomrule
\end{tabular}
\caption{Activities and key concepts}
\label{tab:activities}
\end{table}

Additionally, we created a facilitator-facing guide, including printouts for students and a supporting slide deck. The guide includes a summary of the objective and a step-by-step facilitation sequence for each activity, including discussion questions and instructional talking points. The first three activities feature a ``doing it together'' section, where learners are introduced to the activity as a class, and an ``on your own section,'' in which learners can work individually or in pairs to apply the concepts on their own.

\subsubsection{Activity 1} ``What Comes Next?'' introduces learners to 1st-order Markov chains (bigrams), or single-state predictions, where each subsequent state (or token) depends only on the immediately preceding one. In the activity, learners predict the following animal emoji after a given sequence of emojis, where each emoji is one token and the most recent emoji is the state. Learners count how often each emoji follows the current one and divide by the total number of transitions out of that emoji (e.g., Given the sequence ABAAB, A is followed by B twice out of three times, thus the probability of A → B = 2/3). The final emoji in a sequence is not counted as it does not have a succeeding token. As learners identify patterns based on the most recent emoji, they can build intuition about how systems make predictions based on prior data and are introduced to the concept of tokenization.

\subsubsection{Activity 2} ``Pair of N-imals'' expands on the prior activity by making the state the two most recent tokens (a 2nd-order chain, or trigram model). Learners count how often each emoji follows a given pair of tokens, rather than a single token. This activity challenges learners to consider how more or less context affects model behaviors, affording discussion about the tradeoffs of memory length in predictive systems. 

\subsubsection{Activity 3} ``Sunny, Rainy, Cloudy'' shifts from making single-state predictions (see ``What Comes Next?'') to modeling the probabilities of an entire system using state transition diagrams. State transition diagrams are a visual representation of a system that depicts states as nodes and arrows as transitions, with probabilities indicating how likely the system moves from one state (or token) to another. Using weather emojis, learners complete a state diagram to represent all possible state-to-state relationships, introducing them to think about prediction models as a system. Transitions that never occur in the data are marked with a probability of 0.

\subsubsection{Activity 4} ``Rock, Paper, Scissors'' prompts learners to iteratively generate their own datasets by playing rock, paper, scissors with each other (with each move as a token) to build a 1st-order model of their partner's next move. This activity is designed for learners to think critically about how data quantity and quality shape model behavior, introducing concepts like bias through a competitive game.

\subsubsection{Activity 5} ``Guess the Emoji'' applies key concepts from the previous activities (e.g., probability, datasets, state transition diagrams), prompting learners to design their own emoji datasets and challenging their peers to model the predictive behaviors of their data using state transition diagrams.

\subsection{Participants}
As described in the previous section, we refined the activities across three pilot workshops. This paper focuses on the second iteration, in which we piloted these activities with 22 teenagers (ages 14-15) at a science center in Northeastern U.S. Of these, 19 provided parental consent and assented to participate in research. All data, artifacts, and quotations in this paper come only from these 19 participants. Sixteen of 19 completed an optional demographic survey. Nine teens identified as Black/African-American, four as Asian/Pacific Islander, three as Latinx/Hispanic, and two as White, with two choosing more than one category. Eleven teens identified as male, and five identified as female. All teens reported using LLM-powered systems, with four using them every day, two twice a week, five once a week, two every other week, and three once a month. All names included in this paper are pseudonyms.

\subsection{Data Collection \& Analysis}
For this pilot study, we collected the teens’ initial conception of how LLMs work through a Big Paper activity \citep{10.1145/2485760.2485763}, recordings of their conversations about Markov chains and AI/ML systems during collaborative work, their completed worksheets and artifacts, and recordings of post-activity reflections. Conducting all five activities took around 90 minutes across two days. 

As an experience report from a single pilot study, our analysis was descriptive and qualitative. Two researchers reviewed all completed worksheets and coded each as correct, incorrect with a minor computational mistake (e.g., a miscount or miscalculation), or incorrect with a conceptual mistake. One researcher reviewed the recordings of collaborative work and reflections to find moments where teens described their understandings of LLMs and Markov chains; bringing back findings to the larger team for feedback and review. The study was approved by University of Pennsylvania's IRB.

\section{Findings}
\subsection{Initial Conceptions of LLMs}
\begin{figure}
    \centering
    \includegraphics[width=1\linewidth,alt={two teen drawn illustrations of teens’ ideas about how ChatGPT works, both suggest that ChatGPT summarizes information}]{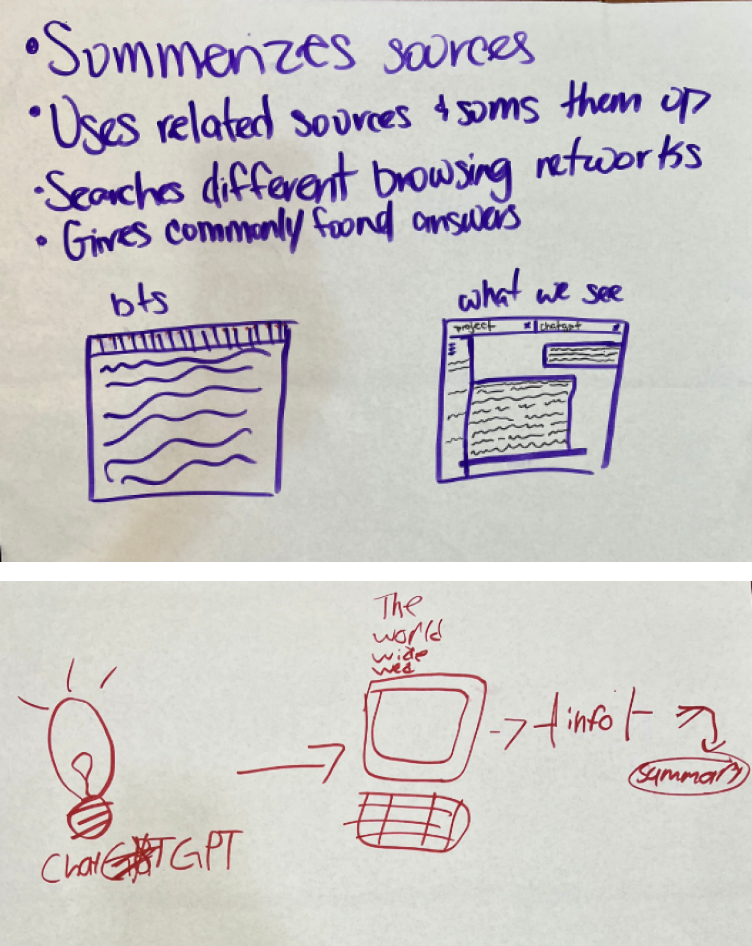}
    \caption{Big Paper activity on how participants think ChatGPT works}
    \label{fig:pre}
\end{figure}

Prior to the unplugged activities, in a Big Paper exercise, where teens worked in separate groups to write and draw how they believed ChatGPT worked, one group of teenagers described the system as one that ``finds sources and summarizes information,'' to which his group nodded along and documented on their paper. Similarly, another learner in a different group described that ChatGPT is a system that ``goes through a bunch of sources at once and finds the answers.'' His group came to an agreement and drew a diagram that showed that the system connected to the World Wide Web, found information, and summarized the information (Figure \ref{fig:pre}). Notably, all groups independently used the language of ``summarizing'' and ``searching,'' an intuitive framing that likely reflected their lived experience interaction with search-driven tools, and which aligned with prior work on young learners’ folk conceptions of AI/ML systems \citep{10.1145/3459990.3460712, 10.1145/3773077.3806107}.

\subsection{Unplugged AI Activities}
\begin{figure}
    \centering
    \includegraphics[width=1\linewidth,alt={Two completed pair of n-imal worksheets, above shows incorrect work and below shows correct work}]{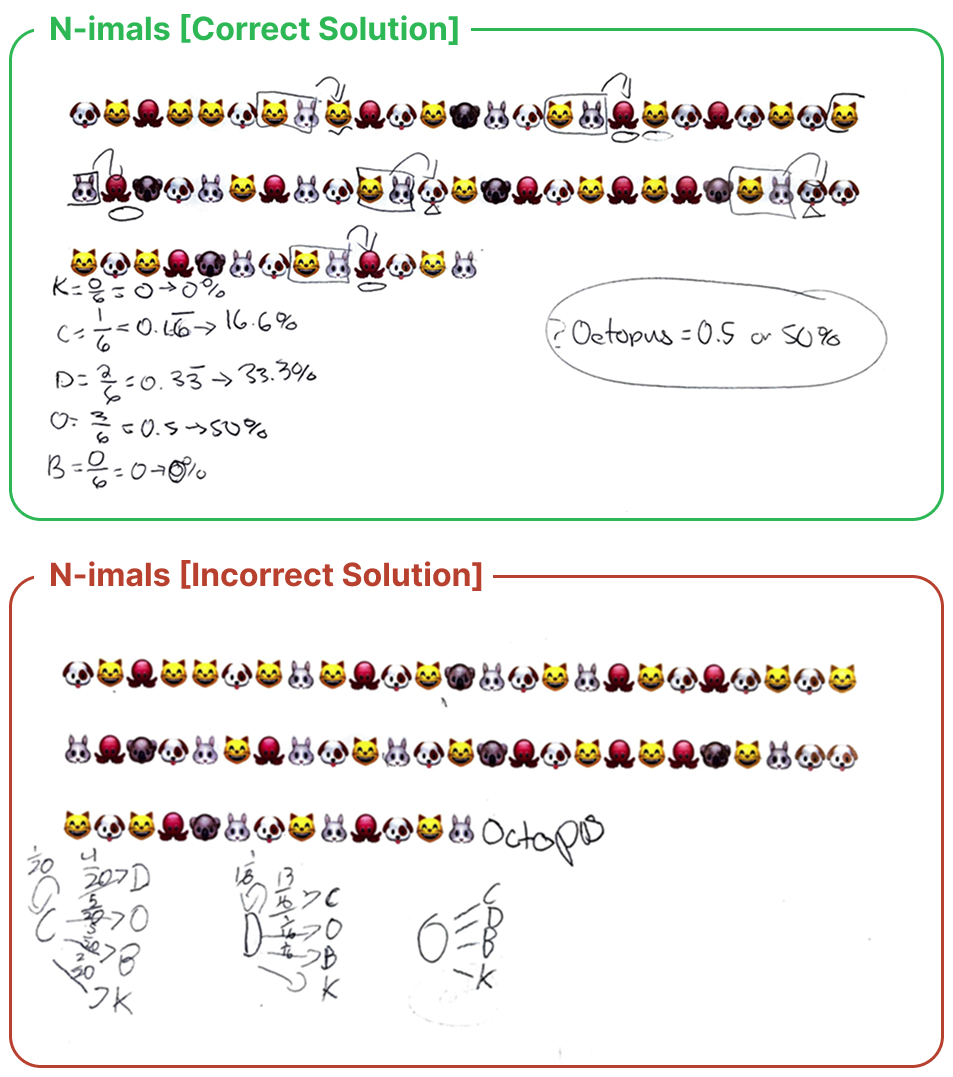}
    \caption{Correct and incorrect Pair of N-imal worksheets}
    \label{fig:nimal}
\end{figure}

All teens successfully completed the first activity (``What Comes Next?''), computing each next-token probability as a fraction with the correct numerator and denominator. Understanding that the activity was based on probability, and only requiring knowledge of fractions, enabled the teens to directly jump into solving the probabilities of a single token. In the second activity (``Pair of N-imals''), however, one pair of learners incorrectly completed the activity (Figure \ref{fig:nimal}), as they continued to compute 1st-order (bigram) Markov chains, where only the last token was considered, when they were prompted to consider the last two. 

\begin{figure}
    \centering
    \includegraphics[width=1\linewidth,alt={Completed worksheet of weather activity, has one error in numerator}]{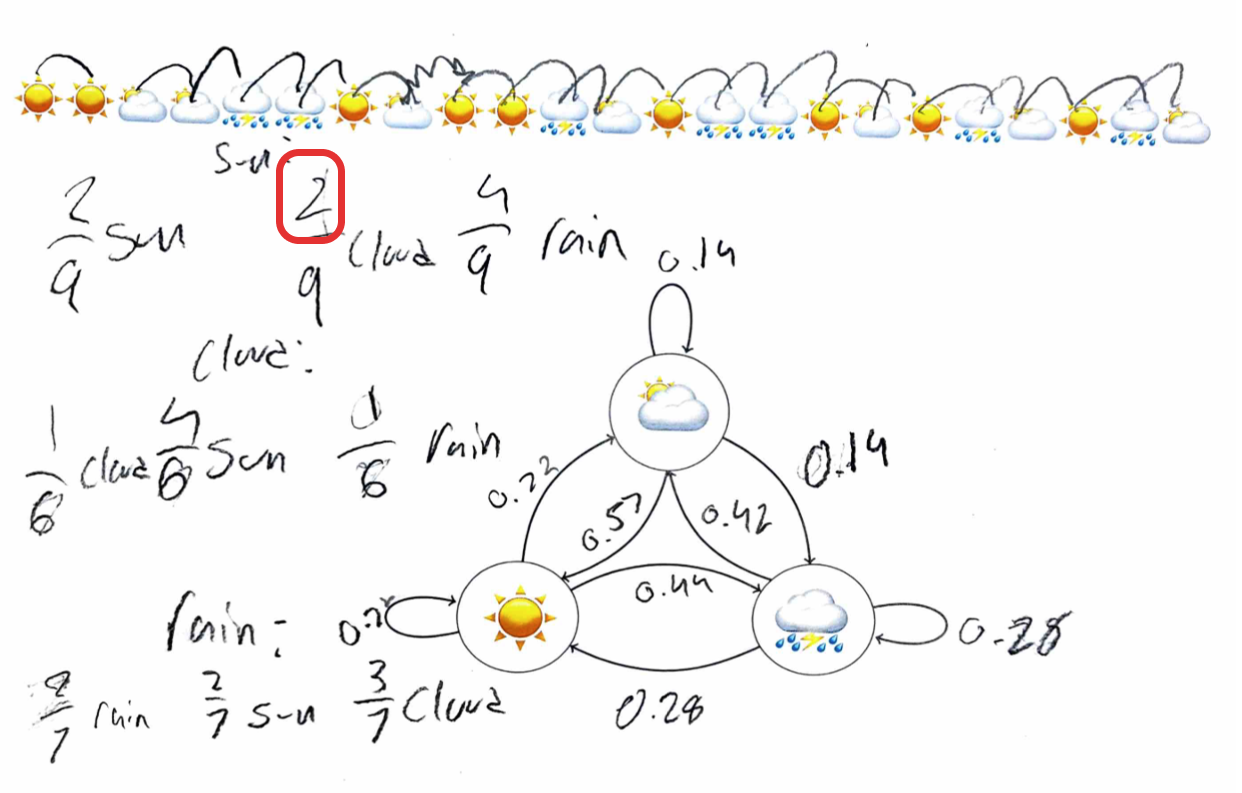}
    \caption{Single mistake in numerator, likely due to a counting error}
    \label{fig:weather}
\end{figure}

Before starting the ``Sunny, Rainy and Cloudy'' activity, one learner aptly remarked, ``When you ask us what the next thing will be, it’s not a definite answer because it’s still up to probability,'' reflecting that she was grasping the core idea that outcomes of Markov chains were nondeterministic, or determined by probability, rather than a fixed outcome. Of nine groups, six groups completed the activity successfully; three pairs made a single error on either one numerator or two connected numerators (Figure \ref{fig:weather}) while getting all denominators correct, which likely can be attributed to errors in counting rather than conceptual understanding. Nonetheless, the facilitator guide was modified to include a tip prompting learners to add the numerators for each weather state and check that they sum to the denominator.

The fourth activity, ``Rock, Paper, Scissors'', surfaced preliminary evidence that learners connected data collection to predictive outcomes during these activities. When a facilitator asked one participant, Trevor, how the activity was going, he described, ``I was cookin'. I caught up to her. I was looking at the probabilities.'' Trevor explained that he had noticed his partner tended to play paper after both paper and rock. In response, he deliberately played scissors more often to skew his own dataset to counter her predicted behavior. In the end, Trevor’s final predicted token beat his partner’s. In a separate group of three (Leo, Avi, and Clara), Leo began coaching Avi on how to skew his own dataset towards a particular move, reflecting an emerging understanding that a model’s output---here, the outcome of the game---is shaped directly by the data collected to produce it. 

\begin{figure}[b]
    \centering
    \includegraphics[width=1\linewidth]{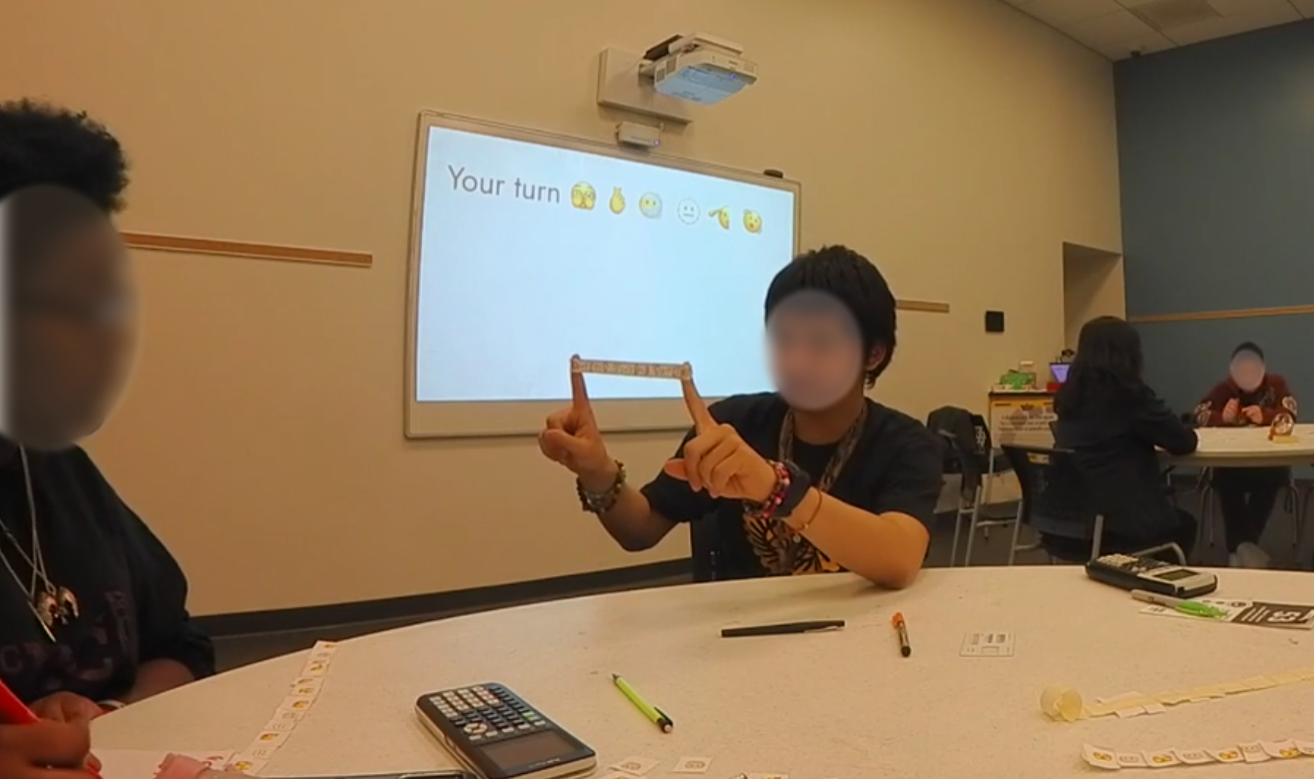}
    \caption{Picture of teens generating a Markov chain using a different group's dataset}
    \label{fig:guess}
\end{figure}

\begin{figure*}[t]
    \centering
    \includegraphics[width=1\linewidth, alt={On the left, learner completed stochastic matrix for Guess the Emoji; right, a state-transition diagram}]{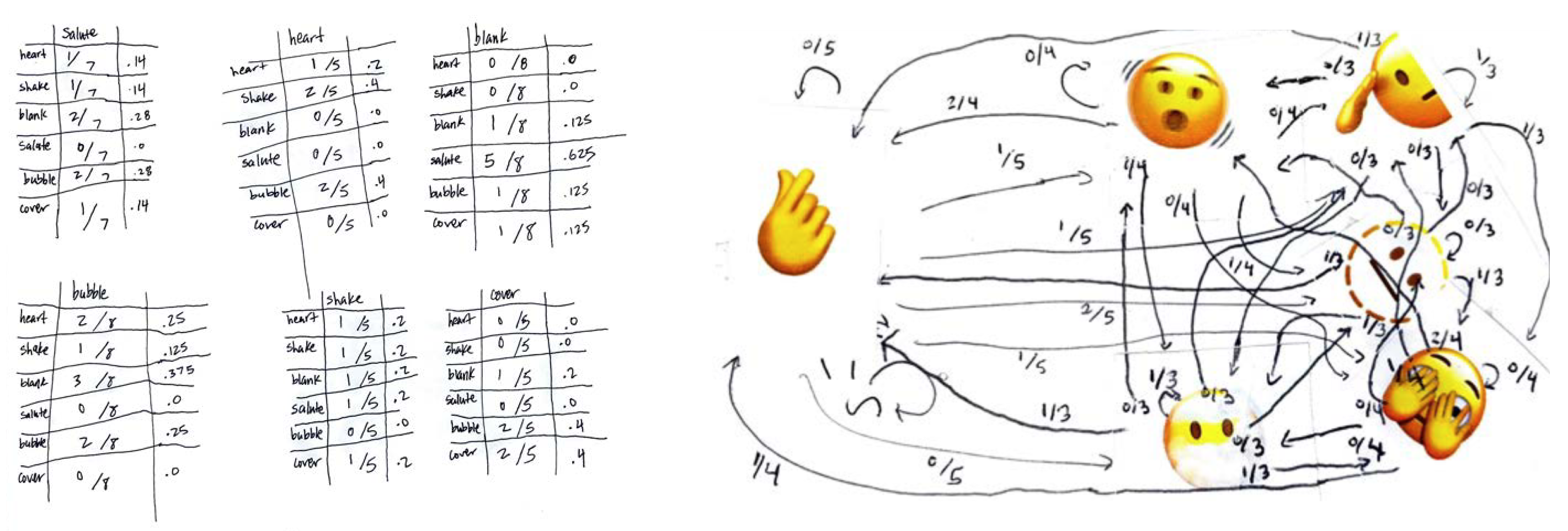}
    \caption{A comprehensive table for all state-to-state transitions (left) and a state-transition diagram (right) for the ``Guess the Emoji'' activity}
    \label{fig:matrix}
\end{figure*}

In the final activity, ``Guess the Emoji,'' (Figure \ref{fig:guess}) teens added more and more emojis to their datasets, reasoning that it would be funny for the group solving their Markov chains to face a longer dataset. Revealing the datasets to one another prompted considerable laughter, but the activity functioned more as additional practice of the concepts from prior activities than for teaching new ones. In addition, even when prompted to just figure out the next most probable token, one learner generated a comprehensive table, akin to a stochastic matrix, with fractions and decimals, for all state-to-state transitions in the dataset (Figure \ref{fig:matrix}), and three other pairs created state transition diagrams. These spontaneous interactions led us to redevelop the final activity as an optional extension that revisits principles from the earlier activities. It also ran longer than intended (20 minutes) and was therefore modified to serve as an extension activity.

\subsection{Post-Activity Reflections}
Post-activity reflections from two participants, Leo and Zach, illustrate how some learners came to describe LLMs over the course of the workshop. In a discussion after the unplugged activities, the facilitators asked what the participants learned about AI from these activities. To this, Leo described that people ``think rationally. [...] AI is not human. It doesn’t know what doesn’t make sense or what does. It just picks words [...] based on probability." Later, after having designed their own generative language models, Leo and Zach both drew explicit connections between how LLMs work and Markov chains:

\begin{quote}
    ``It’s kind of like what we did with the pattern recognition thing (Markov chains) where we used math to determine what would be the most probable word that comes next. It can either do like, do that, or it can take certain sections of one thing and take out another section of another thing to help build a response.''
\end{quote}

These preliminary findings suggest that the activities supported both an understanding of probabilistic models and also provided learners with more concrete vocabulary, such as probability, data, training, or tokens, for reasoning about how generative AI systems more broadly produce outputs.

\section{Discussion}
In the workshop, we used unplugged AI activities to introduce Markov chains to teens. In the following sections, we discuss where we were successful in introducing these concepts to teens, what was challenging, and outline design considerations for future activities.

All workshop activities were engaging and feasible for teens with no prior background in probabilistic modeling. Every group produced a completed worksheet for each activity, and errors were mostly computational rather than conceptual. Our activities complement prior unplugged work by emphasizing explicit calculation. LLMs Unplugged \citep{10.1145/3786228.3786237} prompts learners to select likely next words ``by eye'' from real text, keeping a close connection to natural language, while The Engine Room \citep{10.1145/3773077.3806141} makes sampling tangible through physical mechanisms like drawing words from a bag or spinning a probability wheel. In our activities, learners calculate probabilities themselves and construct transition tables and state transition diagrams, so each prediction can be traced back to the data. Because the first activity, ``What Comes Next?,'' required only a working knowledge of fractions, all groups were able to begin reasoning about single-token predictions immediately, suggesting a low barrier to entry. The fourth activity, ``Rock, Paper, Scissors,'' was particularly engaging, since its competitive, gamified framing motivated learners to evaluate patterns in their data and engage in probabilistic reasoning as a strategy for winning.

Engaging with the inner workings of probabilistic models also drew teens into discussions of the differences between human and artificial intelligence. Whereas teens initially described LLMs as systems that ``search the World Wide Web'' or ``find sources and summarize information,'' some participants' post-activity reflections described generative AI in probabilistic terms; as Leo put it, AI ``picks words based on probability.'' While we cannot attribute this change to the activities across the cohort, these examples suggest the activities gave some learners a concrete, calculable model to reason with.

We also observed a few challenges during our workshop. In ``Pair of N-imals,'' one pair continued to solve for 1st-order (bigram) Markov chains, considering only the most recent token, when the activity prompted them to account for the previous two. It is unclear whether this reflected a misreading of the instructions or a need for additional scaffolding of the n-gram concept; regardless, we modified the printouts to better indicate the final two tokens and scaffold the transition from single- to multi-token context more explicitly.

A broader challenge concerns connections to LLMs and generative AI. The activity design incorporated emojis to make the material entertaining and accessible across age groups, but representing concepts through emojis rather than words may distance learners from the language-based systems they encounter in everyday life. More broadly, introducing Markov chains as a standalone technical concept is an informative first step, but linking these concepts to how generative language models work, and further, how those models affect the teens who are already avid users of these technologies, is critical for fostering learners’ understanding of these systems. Without this, learners may walk away with an isolated skill (e.g., completing a state transition diagram) that is disconnected from their everyday experiences with AI/ML systems. Across the workshop, we found that activities, each targeting a discrete concept within probabilistic modeling, need to be explicitly tied back to the broader topic of why these concepts matter in learners’ own lives. We see the potential of adapting these activities so that they can be incorporated into mathematics, language arts, or computing classes and target subject-specific standards. 

To this point, additional activities should be utilized or designed to transition learners’ understandings of Markov chains to LLMs and other AI/ML systems that they encounter every day. Possible avenues may include the use of the plugged Markov Chain Demo \citep{chen2024markov} or designing additional, text-based unplugged Markov chain activities to bridge the connection between emoji-based Markov chains and language generation. In our pilot, we addressed this by asking teens to select songs they enjoyed and, using the RiTa library\footnote{https://github.com/dhowe/ritajs/} and a provided web-based application, tokenize the lyrics and generate new “song lyrics” via Markov chains, bridging the paper-based activities to the generative AI systems that the learners were more familiar with.

\section{Conclusion}
Unplugged AI activities can offer a low-barrier, hands-on introduction to generative language modeling, providing the foundation for discussions on the functional qualities of generative language models. By predicting, collecting data, identifying patterns, and building paper-based Markov chain models, teenagers can explore multiple facets of core AI/ML concepts while developing capabilities as designers and critical thinkers of algorithmic systems.

\section*{Acknowledgments}
This work was partially supported by the Penn AI Fellowship awarded to Luis Morales-Navarro. We thank Lucianne Servat and Elo Esalomi for support in data collection.

\bibliography{aaai2027}

@inproceedings{10.1145/3702652.3744217,
author = {Gu, Xingjian (Lance) and Ericson, Barbara J.},
title = {AI Literacy in K-12 and Higher Education in the Wake of Generative AI: An Integrative Review},
year = {2025},
isbn = {9798400713408},
publisher = {Association for Computing Machinery},
address = {New York, NY, USA},
url = {https://doi.org/10.1145/3702652.3744217},
doi = {10.1145/3702652.3744217},
booktitle = {Proceedings of the 2025 ACM Conference on International Computing Education Research V.1},
pages = {125–140},
numpages = {16},
location = {
},
series = {ICER '25}
}

@misc{cnn2026denmark,
  author       = {{CNN}},
  title        = {Denmark Takes Measures to Address AI Cheating in Schools},
  year         = {2026},
  month        = aug,
  day          = {7},
  howpublished = {\url{https://edition.cnn.com/2026/08/07/europe/ai-cheating-measures-schools-denmark-intl-scli}},
  note         = {Accessed September 6, 2026}
}

@article{nytimes2026lausd,
  author       = {Goldstein, Dana},
  title        = {Los Angeles Unified School District Bans Artificial Intelligence in Schools},
  journal      = {The New York Times},
  year         = {2026},
  month        = sep,
  day          = {3},
  howpublished = {\url{https://www.nytimes.com/2026/09/03/us/lausd-schools-ban-ai-artificial-intelligence.html}},
  note         = {Accessed September 6, 2026}
}

@misc{nyc2026mamdani,
  author       = {{New York City Mayor's Office}},
  title        = {Mayor Mamdani and Chancellor Samuels Put Students First with Nation's Broadest Generative AI Moratorium in Schools},
  year         = {2026},
  month        = sep,
  day          = {2},
  howpublished = {\url{https://www.nyc.gov/mayors-office/news/2026/09/mayor-mamdani-and-chancellor-samuels-put-students-first-with-nat}},
  note         = {Accessed September 6, 2026}
}

@misc{reuters2026norway,
  author       = {{Reuters}},
  title        = {Norway Imposes Near Ban on AI in Elementary School},
  year         = {2026},
  month        = jun,
  day          = {19},
  howpublished = {\url{https://www.reuters.com/technology/norway-imposes-near-ban-ai-elementary-school-2026-06-19/}},
  note         = {Accessed September 6, 2026}
}

@article{Smyslova02112025,
author = {Daria Smyslova and Shiyan Jiang and Carolyn P. Rosé and Rebecca Ellis and Jie Chao and Qiuqing Li},
title = {Introducing generative AI with Markov Chains: Gendered patterns of competence in English Language Arts classrooms},
journal = {The Journal of Educational Research},
volume = {118},
number = {6},
pages = {701--715},
year = {2025},
publisher = {Routledge},
doi = {10.1080/00220671.2025.2510409},


URL = { 
    
        https://doi.org/10.1080/00220671.2025.2510409
    
    

},
eprint = { 
    
        https://doi.org/10.1080/00220671.2025.2510409
    
    

}

}

@inproceedings{ali2024constructing,
  title={Constructing dreams using generative AI},
  author={Ali, Safinah and Ravi, Prerna and Williams, Randi and DiPaola, Daniella and Breazeal, Cynthia},
  booktitle={Proceedings of the AAAI Conference on Artificial Intelligence},
  volume={38},
  pages={23268--23275},
  year={2024}
}

@misc{bell1998computer,
  title={Computer Science Unplugged: Off-line activities and games for all ages},
  author={Bell, Timothy C and Witten, Ian H and Fellows, Mike},
  year={1998},
  publisher={Computer Science Unplugged Christchurch, New Zealand}
}

@inproceedings{10.1145/3773077.3806141,
author = {Connelly, Luke and Graves Petersen, Marianne and H{\o}holt Kaspersen, Magnus and Musaeus, Line Have and Bilstrup, Karl-Emil Kj{\ae}r},
title = {Addressing the NLP in the Classroom: Empowering High School Teachers through Participatory Design of Hands-on NLP Activities},
year = {2026},
isbn = {9798400722837},
publisher = {Association for Computing Machinery},
address = {New York, NY, USA},
url = {https://doi.org/10.1145/3773077.3806141},
doi = {10.1145/3773077.3806141},
booktitle = {Proceedings of the 25th Annual ACM Interaction Design and Children Conference},
pages = {688–702},
numpages = {15},
location = {
},
series = {IDC '26}
}

@misc{dayIntroChains,
  author       = {Day, Martin V.},
  title        = {A Mathematical Introduction to Markov Chains},
  year         = {2018},
  howpublished = {\url{https://personal.math.vt.edu/day/IntroChains/}},
  note         = {Accessed September 4, 2026}
}

@misc{codeorg2023chatbots,
  author       = {{Code AI}},
  title        = {How AI Works: Chatbots and Large Language Models},
  year         = {2026},
  howpublished = {YouTube video},
  url          = {https://www.youtube.com/watch?v=X-AWdfSFCHQ},
  note         = {Accessed: November 11, 2026}
}

@inproceedings{10.3115/1118108.1118109,
author = {van Halteren, Hans},
title = {Teaching NLP/CL through games: the case of parsing},
year = {2002},
publisher = {Association for Computational Linguistics},
address = {USA},
url = {https://doi.org/10.3115/1118108.1118109},
doi = {10.3115/1118108.1118109},
booktitle = {Proceedings of the ACL-02 Workshop on Effective Tools and Methodologies for Teaching Natural Language Processing and Computational Linguistics - Volume 1},
pages = {1–9},
numpages = {9},
location = {Philadelphia, Pennsylvania},
series = {ETMTNLP '02}
}

@article{10.1145/3465074.3465078,
author = {Long, Duri and Moon, Jonathan and Magerko, Brian},
title = {Unplugged assignments for K-12 AI education},
year = {2021},
issue_date = {March 2021},
publisher = {Association for Computing Machinery},
address = {New York, NY, USA},
volume = {7},
number = {1},
url = {https://doi.org/10.1145/3465074.3465078},
doi = {10.1145/3465074.3465078},
journal = {AI Matters},
month = jul,
pages = {10–12},
numpages = {3}
}

@inproceedings{10.1609/aaai.v40i47.41515,
author = {Reddig, Jennifer M. and Moon, Scott and Crutcher, Kaitlyn and MacLellan, Christopher J.},
title = {AI unplugged: embodied interactions for AI literacy in higher education},
year = {2026},
isbn = {978-1-57735-906-7},
publisher = {AAAI Press},
url = {https://doi.org/10.1609/aaai.v40i47.41515},
doi = {10.1609/aaai.v40i47.41515},
booktitle = {Proceedings of the Fortieth AAAI Conference on Artificial Intelligence and Thirty-Eighth Conference on Innovative Applications of Artificial Intelligence and Sixteenth Symposium on Educational Advances in Artificial Intelligence},
articleno = {4614},
numpages = {9},
series = {AAAI'26/IAAI'26/EAAI'26}
}

@inproceedings{10.1145/3459990.3460712,
author = {Druga, Stefania and Ko, Amy J},
title = {How do children’s perceptions of machine intelligence change when training and coding smart programs?},
year = {2021},
isbn = {9781450384520},
publisher = {Association for Computing Machinery},
address = {New York, NY, USA},
url = {https://doi.org/10.1145/3459990.3460712},
doi = {10.1145/3459990.3460712},
booktitle = {Proceedings of the 20th Annual ACM Interaction Design and Children Conference},
pages = {49–61},
numpages = {13},
location = {Athens, Greece},
series = {IDC '21}
}

@inproceedings{10.1145/3786228.3786237,
author = {Swift, Ben},
title = {LLMs Unplugged: Teaching Resources for a ChatGPT World},
year = {2026},
isbn = {9798400723520},
publisher = {Association for Computing Machinery},
address = {New York, NY, USA},
url = {https://doi.org/10.1145/3786228.3786237},
doi = {10.1145/3786228.3786237},
booktitle = {Proceedings of the 28th Australasian Computing Education Conference},
pages = {55–62},
numpages = {8},
location = {
},
series = {ACE '26}
}

@inproceedings{10.1145/2485760.2485763,
author = {Yip, Jason and Clegg, Tamara and Bonsignore, Elizabeth and Gelderblom, Helene and Rhodes, Emily and Druin, Allison},
title = {Brownies or bags-of-stuff? domain expertise in cooperative inquiry with children},
year = {2013},
isbn = {9781450319188},
publisher = {Association for Computing Machinery},
address = {New York, NY, USA},
url = {https://doi.org/10.1145/2485760.2485763},
doi = {10.1145/2485760.2485763},
booktitle = {Proceedings of the 12th International Conference on Interaction Design and Children},
pages = {201–210},
numpages = {10},
location = {New York, New York, USA},
series = {IDC '13}
}

@inproceedings{10.1145/3392063.3394396,
author = {DiPaola, Daniella and Payne, Blakeley H. and Breazeal, Cynthia},
title = {Decoding design agendas: an ethical design activity for middle school students},
year = {2020},
isbn = {9781450379816},
publisher = {Association for Computing Machinery},
address = {New York, NY, USA},
url = {https://doi.org/10.1145/3392063.3394396},
doi = {10.1145/3392063.3394396},
booktitle = {Proceedings of the Interaction Design and Children Conference},
pages = {1–10},
numpages = {10},
location = {London, United Kingdom},
series = {IDC '20}
}

@InProceedings{pmlr-v207-michaeli23a,
  title = 	 {Data, Trees, and Forests - Decision Tree Learning in K-12 Education},
  author =       {Michaeli, Tilman and Seegerer, Stefan and Kerber, Lennard and Romeike, Ralf},
  booktitle = 	 {Proceedings of the Third Teaching Machine Learning and Artificial Intelligence Workshop},
  pages = 	 {37--41},
  year = 	 {2023},
  editor = 	 {Kinnaird, Katherine M. and Steinbach, Peter and Guhr, Oliver},
  volume = 	 {207},
  series = 	 {Proceedings of Machine Learning Research},
  month = 	 {19--23 Sep},
  publisher =    {PMLR},
  url = 	 {https://proceedings.mlr.press/v207/michaeli23a.html}
}

@inproceedings{10.1145/3545945.3569772,
author = {Ma, Ruizhe and Sanusi, Ismaila Temitayo and Mahipal, Vaishali and Gonzales, Joseph E. and Martin, Fred G.},
title = {Developing Machine Learning Algorithm Literacy with Novel Plugged and Unplugged Approaches},
year = {2023},
isbn = {9781450394314},
publisher = {Association for Computing Machinery},
address = {New York, NY, USA},
url = {https://doi.org/10.1145/3545945.3569772},
doi = {10.1145/3545945.3569772},
booktitle = {Proceedings of the 54th ACM Technical Symposium on Computer Science Education V. 1},
pages = {298–304},
numpages = {7},
location = {Toronto ON, Canada},
series = {SIGCSE 2023}
}

@inproceedings{10.1145/3706599.3719876,
author = {Darabipourshiraz, Hasti and Bhat, Maalvika and Long, Duri},
title = {Introducing AI Without Computers: Hands-On Literacy and Ethical Sense-Making for Young Learners},
year = {2025},
isbn = {9798400713958},
publisher = {Association for Computing Machinery},
address = {New York, NY, USA},
url = {https://doi.org/10.1145/3706599.3719876},
doi = {10.1145/3706599.3719876},
booktitle = {Proceedings of the Extended Abstracts of the CHI Conference on Human Factors in Computing Systems},
articleno = {353},
numpages = {11},
location = {
},
series = {CHI EA '25}
}

@inproceedings{baldoni2026unplugged,
  title={Unplugged activities on machine learning and their evaluation through mental states attribution},
  author={Baldoni, Matteo and Baroglio, Cristina and Bucciarelli, Monica and Capecchi, Sara and Castellani, Leonardo and Gandolfi, Elena and Ian{\`\i}, Francesco and Marengo, Elisa and Micalizio, Roberto},
  booktitle={Proceedings of the AAAI Conference on Artificial Intelligence},
  volume={40},
  pages={40555--40563},
  year={2026}
}

@inproceedings{10.1145/1539024.1508951,
author = {Nishida, Tomohiro and Kanemune, Susumu and Idosaka, Yukio and Namiki, Mitaro and Bell, Tim and Kuno, Yasushi},
title = {A CS unplugged design pattern},
year = {2009},
isbn = {9781605581835},
publisher = {Association for Computing Machinery},
address = {New York, NY, USA},
url = {https://doi.org/10.1145/1508865.1508951},
doi = {10.1145/1508865.1508951},
booktitle = {Proceedings of the 40th ACM Technical Symposium on Computer Science Education},
pages = {231–235},
numpages = {5},
location = {Chattanooga, TN, USA},
series = {SIGCSE '09}
}

@article{bell2009computer,
  title={Computer science unplugged: School students doing real computing without computers},
  author={Bell, Tim and Alexander, Jason and Freeman, Isaac and Grimley, Mick},
  journal={New Zealand Journal of applied computing and information technology},
  volume={13},
  number={1},
  pages={20--29},
  year={2009}
}

@misc{chen2024markov,
  author       = {Chen, Jessica and Dewan, Aditya and Touretzky, David S.},
  title        = {Markov Chain Demo: Activity Guide},
  year         = {2024},
  month        = nov,
  day          = {1},
  howpublished = {AI4K12},
  url          = {https://ai4k12.org/wp-content/uploads/markov_chain-activity_guide.pdf}
}

@inproceedings{10.1145/3737609.3747112,
author = {Connelly, Luke and Bilstrup, Karl-Emil Kj{\ae}r and Petersen, Marianne Graves},
title = {Beyond LLMs as Black Boxes: Activities and an Educational Tool Supporting Unplugged and Digital AI Learning Activities for K-12 Classrooms},
year = {2025},
isbn = {9798400719684},
publisher = {Association for Computing Machinery},
address = {New York, NY, USA},
url = {https://doi.org/10.1145/3737609.3747112},
doi = {10.1145/3737609.3747112},
booktitle = {Adjunct Proceedings of the Sixth Decennial Aarhus Conference: Computing X Crisis},
articleno = {11},
numpages = {4},
location = {
},
series = {AAR Adjunct '25}
}

@inproceedings{10.1145/3769994.3770061,
author = {Pope, Nicolas and Tedre, Matti},
title = {A Teachable Machine for Transformers},
year = {2025},
isbn = {9798400715990},
publisher = {Association for Computing Machinery},
address = {New York, NY, USA},
url = {https://doi.org/10.1145/3769994.3770061},
doi = {10.1145/3769994.3770061},
booktitle = {Proceedings of the 25th Koli Calling International Conference on Computing Education Research},
articleno = {50},
numpages = {3},
location = {
},
series = {Koli Calling '25}
}

@article{10.1145/3786761,
author = {Darabipourshiraz, Hasti and Murakami Ng, Lily and Wang, Grace and Rollins, Sophie and Long, Duri},
title = {AI Unplugged: Exploring Pathways from Physical Simulation to Conceptualization of AI Reasoning Processes},
year = {2026},
issue_date = {September 2026},
publisher = {Association for Computing Machinery},
address = {New York, NY, USA},
volume = {26},
number = {3},
url = {https://doi.org/10.1145/3786761},
doi = {10.1145/3786761},
journal = {ACM Trans. Comput. Educ.},
month = apr,
articleno = {41},
numpages = {37}
}

@inproceedings{10.1007/978-3-030-33759-9_10,
author = {Lindner, Annabel and Seegerer, Stefan and Romeike, Ralf},
title = {Unplugged Activities in the Context of AI},
year = {2019},
isbn = {978-3-030-33758-2},
publisher = {Springer-Verlag},
address = {Berlin, Heidelberg},
url = {https://doi.org/10.1007/978-3-030-33759-9_10},
doi = {10.1007/978-3-030-33759-9_10},
booktitle = {Informatics in Schools. New Ideas in School Informatics: 12th International Conference on Informatics in Schools: Situation, Evolution, and Perspectives, ISSEP 2019, Larnaca, Cyprus, November 18–20, 2019, Proceedings},
pages = {123–135},
numpages = {13},
location = {Larnaca, Cyprus}
}

@inproceedings{10.1145/3773077.3812183,
author = {Klemettil{\"a}, Pauli A. and Sharma, Sumita and Iivari, Netta and Vent{\"a}-Olkkonen, Leena and Hartikainen, Heidi and Rajanen, Mikko and Holappa, Jenni and Morales-Navarro, Luis and Kafai, Yasmin},
title = {AI Literacy with Teenagers: Exploring Large Language Models Beneath the Surface},
year = {2026},
isbn = {9798400722837},
publisher = {Association for Computing Machinery},
address = {New York, NY, USA},
url = {https://doi.org/10.1145/3773077.3812183},
doi = {10.1145/3773077.3812183},
booktitle = {Proceedings of the 25th Annual ACM Interaction Design and Children Conference},
pages = {1251–1256},
numpages = {6},
location = {
},
series = {IDC '26}
}

@inproceedings{10.1145/3773077.3806107,
author = {Morales-Navarro, Luis and Noh, Daniel J. and Servat, Lucianne and Netting, Carly and Kafai, Yasmin and Metaxa, Dana{\'e}},
title = {Building to Understand: Examining Teens’ Technical and Socio-Ethical Pieces of Understanding in the Construction of Small Generative Language Models},
year = {2026},
isbn = {9798400722837},
publisher = {Association for Computing Machinery},
address = {New York, NY, USA},
url = {https://doi.org/10.1145/3773077.3806107},
doi = {10.1145/3773077.3806107},
booktitle = {Proceedings of the 25th Annual ACM Interaction Design and Children Conference},
pages = {262–280},
numpages = {19},
location = {
},
series = {IDC '26}
}

@article{10.1145/3641552,
author = {Tseng, Tiffany and Davidson, Matt J. and Morales-Navarro, Luis and Chen, Jennifer King and Delaney, Victoria and Leibowitz, Mark and Beason, Jazbo and Shapiro, R. Benjamin},
title = {Co-ML: Collaborative Machine Learning Model Building for Developing Dataset Design Practices},
year = {2024},
issue_date = {June 2024},
publisher = {Association for Computing Machinery},
address = {New York, NY, USA},
volume = {24},
number = {2},
url = {https://doi.org/10.1145/3641552},
doi = {10.1145/3641552},
journal = {ACM Trans. Comput. Educ.},
month = apr,
articleno = {25},
numpages = {37}
}


\end{document}